\documentclass[11pt]{article}
\usepackage{amsfonts}
\usepackage{epsf}
\def\preprint#1{%
    \thispagestyle{empty}~\newline\vspace*{-22.65mm}  
    \begin{flushright}  
    \begin{tabular}{l} #1 \end{tabular}  
    \end{flushright}  
    \vspace{1cm}}  
\def\lsim{\mathrel{\lower2.5pt\vbox{\lineskip=0pt\baselineskip=0pt  
          \hbox{$<$}\hbox{$\sim$}}}}  
\def\gsim{\mathrel{\lower2.5pt\vbox{\lineskip=0pt\baselineskip=0pt  
          \hbox{$>$}\hbox{$\sim$}}}}  
\def\real{\mathrel{\lower.0pt \hbox{$I\!\!R$}}}  
\renewcommand{\theequation}{\arabic{section}.\arabic{equation}}  
\begin{document}    
\markright{Cosmic Acceleration and Modified Gravitational Models \hfil} 
\title{{\large \preprint{{\tt astro-ph/0411209}}} \huge
Cosmic Acceleration and Modified Gravitational Models \thanks{Based on talks given
at MRST-04, PASCOS-04 and COSMO-04.} \vspace{.8cm}}

\author{\Large Damien A. Easson \thanks{easson@physics.syr.edu}\\ [2mm]  
{\it
Department of Physics, Syracuse University} \\
{\it
Syracuse, NY 13244-1130, USA}}\date{{\today}}  
\maketitle  
\begin{abstract}
\large  
\baselineskip .7cm 

\noindent
There is now overwhelming observational evidence that our Universe is accelerating in its expansion. I discuss how
modified gravitational models can provide an explanation for this observed late-time cosmic acceleration. We consider specific
low-curvature corrections to the Einstein-Hilbert action. Many of these models generically contain unstable de Sitter
solutions and, depending on the parameters of the theory, late-time accelerating attractor solutions.   
\end{abstract}  
\thispagestyle{empty}
\vfill  
\def\ie{{\em i.e.\,}}  
\def\eg{{\em e.g.\,}}  
\def\etc{{\em etc.\,}}  
\def\etal{{\em et al.\,}}  
\def\mcS{{\mathcal S}}  
\def\I{{\mathcal I}}  
\def\mL{{\mathcal L}}  
\def\H{{\mathcal H}}  
\def\M{{\mathcal M}}  
\def\N{{\mathcal N}} 
\def\O{{\mathcal O}} 
\def\T{{\mathcal T}} 
\def\cP{{\mathcal P}} 
\def\R{{\mathcal R}}  
\def\K{{\mathcal K}}  
\def\W{{\mathcal W}} 
\def\mM{{\mathcal M}} 
\def\mJ{{\mathcal J}} 
\def\mP{{\mathbf P}} 
\def\mT{{\mathbf T}} 
\def\mR{{\mathbf R}}
\def\mS{{\mathbf S}}
\def\mX{{\mathbf X}}
\def\mZ{{\mathbf Z}}
\def\eff{{\mathrm{eff}}}  
\def\Newton{{\mathrm{Newton}}}  
\def\bulk{{\mathrm{bulk}}}  
\def\brane{{\mathrm{brane}}}  
\def\matter{{\mathrm{matter}}}  
\def\tr{{\mathrm{tr}}}  
\def\normal{{\mathrm{normal}}}  
\def\implies{\Rightarrow}  
\def\half{{1\over2}}  
\newcommand{\da}{\dot{a}}
\newcommand{\db}{\dot{b}}
\newcommand{\dn}{\dot{n}}
\newcommand{\dda}{\ddot{a}}
\newcommand{\ddb}{\ddot{b}}
\newcommand{\ddn}{\ddot{n}}
\def\be{\begin{equation}}
\def\ee{\end{equation}}
\def\bea{\begin{eqnarray}}
\def\eea{\end{eqnarray}}
\def\bs{\begin{subequations}}
\def\es{\end{subequations}}
\def\g{\gamma}
\def\G{\Gamma}
\def\vp{\varphi}
\def\mpl{M_{\rm P}}
\def\ms{M_{\rm s}}
\def\ls{\ell_{\rm s}}
\def\lmin{\ell_{\rm min}}
\def\lp{\ell_{\rm pl}}
\def\l{\lambda}
\def\gs{g_{\rm s}}
\def\d{\partial}
\def\co{{\cal O}}
\def\sp{\;\;\;,\;\;\;}
\def\spa{\;\;\;}
\def\r{\rho}
\def\dr{\dot r}
\def\dt{\dot\varphi}
\def\e{\epsilon}
\def\k{\kappa}
\def\m{\mu}
\def\n{\nu}
\def\om{\omega}
\def\tn{\tilde \nu}
\def\p{\phi}
\def\vp{\varphi}
\def\r{\rho}
\def\s{\sigma}
\def\t{\tau}
\def\x{\chi}
\def\z{\zeta}
\def\a{\alpha}
\def\de{\delta}
\def\bra#1{\left\langle #1\right|}
\def\ket#1{\left| #1\right\rangle}
\newcommand{\stt}{\small\tt}
\renewcommand{\theequation}{\arabic{section}.\arabic{equation}}
\newcommand{\eq}[1]{equation~(\ref{#1})}
\newcommand{\eqs}[2]{equations~(\ref{#1}) and~(\ref{#2})}
\newcommand{\eqto}[2]{equations~(\ref{#1}) to~(\ref{#2})}
\newcommand{\GeV}{\mbox{GeV}}
\def\ricci{R_{\m\n} R^{\m\n}}
\def\riemann{R_{\m\n\l\s} R^{\m\n\l\s}}
\def\SIZE{1.00} 
\baselineskip .7cm   

\section{Introduction}
One of the most profound discoveries of observational physics is that the universe is accelerating in its expansion. Evidence
for this acceleration is provided by recent supernovae data together with measured CMB anisotropies and large 
scale structure data \cite{Riess:1998cb}.
Despite a large number of attempts to explain this striking phenomenon (for example, \emph{dark energy} associated to some 
new scalar field~\cite{Wetterich:fm} or a cosmological constant~\cite{Carroll:2000fy}) all current models require at least some degree of fine-tuning and are incomplete.
Due to the lack of a satisfactory explanation, it seems reasonable to consider the possibility that the observed 
acceleration is an indication that General Relativity requires some sort of low-energy
modification~\cite{Deffayet:2001pu,Carroll:2003wy}. This radical possibility motivated the recent work of Carroll, Duvvuri, Trodden and Turner 
(CDTT)~\cite{Carroll:2003wy}. These authors considered a simple modification to the Einstein-Hilbert action of the form $R^{n}$, with $n<0$.
They showed that under certain circumstances such models can lead to late-time acceleration of the expanding universe.
In these notes, I review the basic principles of the CDTT model and explain how these principles can be applied to more general modified 
graviational models~\cite{Carroll:2004de}.

\section{Modifications of General Relativity}
The cosmic evolution of our Universe is believed to be described by the Einstein-Hilbert action (coupled to matter)
\be
S=\frac{\mpl^{2}}{2}\int d^{4}\!x \, \sqrt{-g}\, R + \int d^{4}\!x \, \sqrt{-g}\,\mL_{m}
\,,
\label{seh}
\ee
where $\mpl \equiv (8 \pi G)^{-1/2}$ is the reduced Planck mass, $R$ is the Ricci Scalar constructed from the metric tensor $g_{\m\n}$ and
$\mL_{m}$ represents some matter Lagrangian. Let us consider modifications of (\ref{seh}), of the general form
\be
S=\frac{\mpl^{2}}{2}\int d^{4}\!x \, \sqrt{-g}\, F(R,\,\ricci ,\, \riemann , \dots) + \int d^{4}\!x \, \sqrt{-g}\,\mL_{m}
\,,
\label{semod}
\ee
where $R_{\m\n}$ and $R_{\m\n\l\s}$ are the Ricci and Riemann tensors, respectively. In the literature, such modifications have been used to
obtain early-time inflation (for example, Starobinsky considered $F(R)=R + R^{2}+$\it conformal anomaly \rm in~\cite{Starobinsky:te}) and 
to try to eliminate curvature singularities in cosmological and black hole spacetimes~\cite{markov}. 
Unlike these high curvature examples, in order to explain late-time 
cosmic acceleration, we are 
interested in modifications that become important at \emph{low curvatures}. We will see that such modified models can lead to late-time 
accelerating vacuum solutions, providing a \emph{purely gravitational} alternative to dark energy.
\subsection{The CDTT Model}
\begin{figure}[ht]
\centerline{\epsfxsize=4.0in\epsfbox{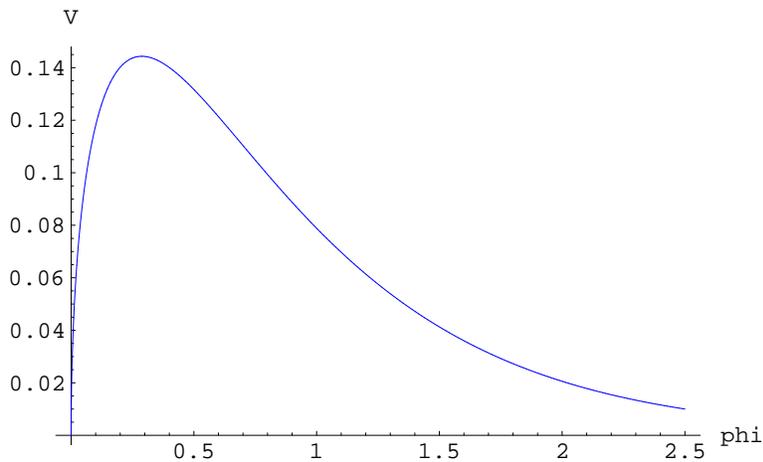}}   
\caption{Plot of the potential $V(\vp)$ in Eq.~(\ref{vofp}). 
\label{fig1}}
\end{figure}

The first examples of these low-curvature modifications where presented by CDTT and involved simple inverse-powers of
the Ricci Scalar (\ie $F(R) =R+m R^{n}$, where $n<0$ and $m$ is some new scale in the theory)~\cite{Carroll:2003wy}. By considering
only inverse powers of $R$, we exclude Minkowski space as a solution and ensure that the correction term dominates the action only at
low curvatures.
As a toy model, consider the case when $n=-1$,
\be
S=\frac{\mpl^{2}}{2}\int d^{4}\!x \, \sqrt{-g}\, \left( R - \frac{\m^{4}}{R}\right) + \int d^{4}\!x \, \sqrt{-g}\,\mL_{m}
\,.
\label{scdtt}
\ee
Here $\mu$ is a new scale in the theory with dimensions of mass. Assuming a flat, Friedmann-Robertson-Walker (FRW) metric:
\be
ds^{2}=-dt^{2}+a^{2}(t)d { \b{x}}^{2}
\,,
\ee
the modified Friedmann equation (in terms of the Hubble parameter $H\equiv \dot{a}/a$) is
\begin{equation}
\label{newfriedmann}
3H^2 - \frac{\mu^4}{12({\dot H}+2H^2)^3}\left(2H{\ddot H}
+15H^2{\dot H}+2{\dot H}^2+6H^4\right) = \frac{\rho_m}{\mpl^2}\,
\end{equation}
where $\rho_{m}$ is the matter energy density and we have assumed a perfect-fluid energy-momentum tensor
\be
T_{\m\n} = (\rho_{m} + P_{m})u_{\m}u_{\n} + P_{m}g_{\m\n}
\,.
\ee
It is a trivial task to map this theory to an Einstein frame with Einstein-Hilbert action minimally coupled to a scalar field $\vp$ with
potential
\begin{equation} 
V(\vp)=\mu^2 \mpl^2
\exp\left(-2\sqrt{\frac{2}{3}}\frac{\vp}{\mpl} \right)\sqrt{\exp
\left(\sqrt{\frac{2}{3}}\frac{\varphi}{\mpl} \right)-1} \,.
\label{vofp}
\end{equation} 
This potential is plotted in Fig.~(\ref{fig1}).
In this model there is a singularity corresponding to $\vp=0$.
From the plot of $V(\vp)$ it is clear that this model exhibits three different possible cosmological behaviors, depending
on the initial conditions for the field $\vp_{i}$ and its velocity $\vp'_{i}$ (where the prime denotes differentiation with respect to the cosmic time coordinate
of the Einstein frame). The field $\vp$ can either roll back down the potential toward a future singularity at $\vp=0$, come to rest
at the top of the potential (corresponding to an unstable de Sitter phase) or roll off to infinity leading to late-time, power-law 
cosmic acceleration with equation of state parameter $w_{eff}=-2/3$. These basic features are not affected by the addition of matter.
By choosing $\mu \simeq 10^{-33} eV$, the modifications to Einstein gravity only become important recently, making this theory a candidate
to explain the observed acceleration of the Universe. It is important to point out this model is different from other models
with scalar fields that attempt to explain dark energy, because our field is \it non-\rm minimally coupled to matter. If the matter
is added directly in the gravity frame (\ref{scdtt}), the Einstein-frame matter density is related to $\rho$ by
\be
\tilde\rho = \exp\left(-2\sqrt{\frac{2}{3}}\frac{\vp}{\mpl} \right) \rho
\,.
\ee

This simple model is easily generalized to corrections of the form $-\mu^{2(n+1)}/R^{n}$. These generalizations maintain the desirable
late-time accelerating solutions with behavior analogous to dark energy with equation of state parameter
\be
w_{eff}=-1 + \frac{2(n+2)}{3(2n+1)(n+1)}
\,.
\ee
It is, therefore, easy to construct models of this type that obey the observational constraints on the equation of state 
parameter $-1.45 < w < -.74$ (at the 95\% confidence level)~\cite{Melchiorri:2002ux}. These simple models provide a new approach to understanding
the dark energy problem. 

\subsubsection{Solar System Constraints}
$F(R)$ theories can have difficulties satisfying certain observational constraints
(for example, solar system constraints)~\cite{chiba}. A simple way to understand this is the following. By introducing a new scalar
field $\phi$, the action (\ref{semod}) with $F$ a function only of $R$, is equivalent to 
\be
S=\frac{\mpl^{2}}{2}\int d^{4}\!x \, \sqrt{-g}\, \left( F(\phi) + F'(\phi) (R-\phi)\right)+ \int d^{4}\!x \, \sqrt{-g}\,\mL_{m}
\,,
\label{actsasha}
\ee
where prime denotes differentiation with respect to $\phi$. The absence of a kinetic term for $\phi$ in (\ref{actsasha}), means that this theory
can be written as a Brans-Dicke type theory, which is of the general form:
\be
S \simeq \int d^{4}\!x \, \sqrt{- \bar g}\, \left(\Phi R - \frac{\omega}{\Phi} \partial_{\mu} \Phi \partial^{\mu}\Phi - U(\Phi)\right)+ S_{m}
\,,
\label{actbd}
\ee
with Brans-Dicke parameter $\omega = 0$. If the scalar field is very light (as in the CDTT model) there are constraints on $\omega$ from 
solar system tests that give $\omega>40,000$~\cite{Bertotti:2003rm}.\footnote{Even the CDTT model is not ruled out by this argument if we live in a place 
where the scalar field is high enough up on the potential in Fig. (\ref{fig1}). Furthermore, some possible ways 
of resolving these instabilities in general are discussed in~\cite{Nojiri:2003ft}.}
\section{Generalized Modified Gravity Models}
As we have seen, the ability to map the $F(R)$ theory to an Einstein frame makes interpretation of solutions particularly
simple. In these notes we would like to consider more complicated modifications such as $F(R,\,\ricci ,\, \riemann , \dots)$. It is not always
possible to map these general models to a familiar Einstein frame and we must, therefore, develop new methods for studying such systems. A specific technique
is presented in our recent paper~\cite{Carroll:2004de}. 

Here we discuss a few simple examples of more general modified gravitational models. To begin we search for general features of such models. Consider the action
\be
S=\int \, d^{4}\!x \, \left(R + f(R,P,Q) \right) +  \int d^{4}\!x \, \sqrt{-g}\,\mL_{m}
\,,
\label{sgmg}
\ee
with $P\equiv \ricci$ and $Q \equiv \riemann$. By defining $f_{R}\equiv \partial f/ \partial R$,
$f_{P}\equiv \partial f/ \partial P$ and $f_{Q}\equiv \partial f/ \partial Q$, the equations of motion derived from (\ref{sgmg})
are
\begin{eqnarray}
R_{\mu\nu} &-& \half\,g_{\mu\nu}\,R-\half\,g_{\mu\nu}\,f \nonumber \\
&+& f_R\,R_{\mu\nu}+2f_P\,R^\alpha{}_\mu\,R_{\alpha\nu}
+2f_Q\,R_{\alpha\beta\gamma\mu}\,R^{\alpha\beta\gamma}{}_\nu \nonumber\\
&+& g_{\mu\nu}\,\Box f_R -\nabla_\mu\nabla_\nu f_R
-2\nabla_\alpha\nabla_\beta[f_P\,R^\alpha{}_{(\mu}{}\delta^\beta{}_{\nu)}]
+\Box(f_P\,R_{\mu\nu}) \nonumber\\
&+& g_{\mu\nu}\,\nabla_\alpha\nabla_\beta(f_P\,R^{\alpha\beta})
-4\nabla_\alpha\nabla_\beta[f_Q\,R^\alpha{}_{(\mu\nu)}{}^\beta]
=8\pi G\,T_{\mu\nu}\ .\label{equaz}
\end{eqnarray}
Let us focus on vacuum solutions to the above field equations. This is physically well motivate since we are interested in studying the new
cosmological features induced by the purely gravitational sector of the theory. 
Mathematically, this provides us with valuable insight into the structure of the equations.

To ensure that Minkowski space is not an allowed solution we consider modifications involving inverse powers of curvature invariants.
It is likely that the addition of such terms will introduce ghost degrees of freedom into the theory.
If ghosts arise we shall require that some unknown mechanism (for example, extra-dimensional effects) cut off the
theory in such a way that the associated instabilities do not appear on cosmological time scales.

The first general feature of the vacuum solutions of (\ref{equaz}), is the existence of constant curvature solutions. 
To see this, we take the trace of~(\ref{equaz}) and substitute $Q=R^2/4$ and $P=R^2/6$ 
(which are identities satisfied by constant curvature spacetimes) into the resulting equation to obtain the algebraic equation:
\begin{equation}
\label{constcurvexp}
\left(2f_{Q}+3f_{P}\right)R^2+6\left(f_R -1\right)R -12f=0 \ .
\end{equation}
Solving this equation for the Ricci scalar yields the constant curvature (de Sitter) vacuum solutions with $R\neq 0$.
This is a very general property of actions of the form~(\ref{sgmg}). 
However, an equally generic feature of such models is that this de Sitter solution is unstable. 
In the CDTT model, the instability is to an accelerating power-law attractor. 
We will see a similar trend in many of the more general models under consideration here.

For simplicity, we now specialize to a class of actions with
\begin{equation}
f(R,P,Q)=-\frac{\mu^{4n+2}}{(aR^2+b P+c Q)^n} \ ,
\label{f_R}
\end{equation}
where $n$ is a positive integer, $\mu$ has dimensions of mass and $a$, $b$ and $c$ are dimensionless constants. In this
case, there are power-law attractors with the following (real) exponents
\begin{equation}
v^{\rm gen}_{1,2}=\frac{8n^2+10n+2-3\alpha\pm
\sqrt\Gamma}{4(n+1)}\ ,
\end{equation}
where
\begin{equation}
\Gamma=9n^2\alpha^2-(80n^3+116n^2+40n+4)\,\alpha
+64n^4+160n^3+132n^2+40n+4 \ .
\end{equation}

Two special cases are
\begin{eqnarray}
f &=& -\frac{m^{4n+2}}{P^n} \ , \nonumber \\
f &=& -\frac{M^{4n+2}}{Q^n} \ ,
\end{eqnarray}
for which the power law attractors, $v_{1,2}$ are
\begin{eqnarray}
v^{(P)}_{1,2}&=&\frac{12n^2+9n+3\pm
\sqrt{144n^2+120n^3-15n^2-30n-3}}{2(3+3n)}
\label{VPN}\\
v^{(Q)}_{1,2}&=&\frac{4n^2+2n+1\pm
\sqrt{16n^4-16n^2-10n-1}}{2n+2}\ .
\label{VQN}
\end{eqnarray}
Having discovered some basic common features of these general models (de Sitter repeller and power-law attractor solutions), 
let us consider some specific examples in greater detail.
\subsection{Inverse powers of R}
\begin{figure}[ht]
\centerline{\epsfxsize=3.5in\epsfbox{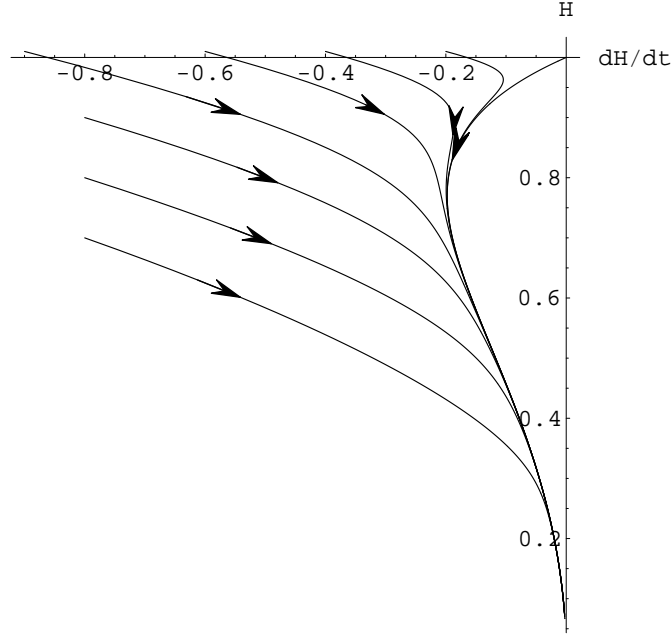}}   
\caption{Phase portrait for the CDTT model with $f(R)=1/R$. The unstable de Sitter solution is at the point $(0,1)$. Solutions
that are initially not accelerating ($\ddot a<0$) evolve toward the power-law, accelerating attractor (with $\ddot a>0$).
\label{fig2}}
\end{figure}

For the simplest case, when $f(R)=-\mu^4/R$, the model is simply that of CDTT with modified Friedmann equation given by
(\ref{newfriedmann}). One way to analyze this system (without going to an Einstein frame) is to simply plot the 
phase space of solutions (see Fig.(\ref{fig2})).
Recall, a generic feature of the models considered here is an unstable de Sitter solution (obtained from Eq.~(\ref{constcurvexp})). 
For the above case of $f(R)=1/R$, the de Sitter solution
is located at the point $(0,1)$ in the $(\dot H, H)$-phase plane depicted in Fig.(\ref{fig2}). The late time power law attractor
has $a(t)\propto t^{p}$, $H=p/t$ and $\dot H=-p/t^{2}$. 
Hence, such attractor solutions appear as curves $H=\sqrt{-p \dot H}$, ($\dot H<0$) in the phase portrait. In the case $1/R$, there is only one
late-time power-law attractor with $p=2$. Following this example, we proceed to examine more complicated modified theories.
\subsection{Inverse powers of $P\equiv \ricci$}
\begin{figure}[ht]
\centerline{\epsfxsize=3.5in\epsfbox{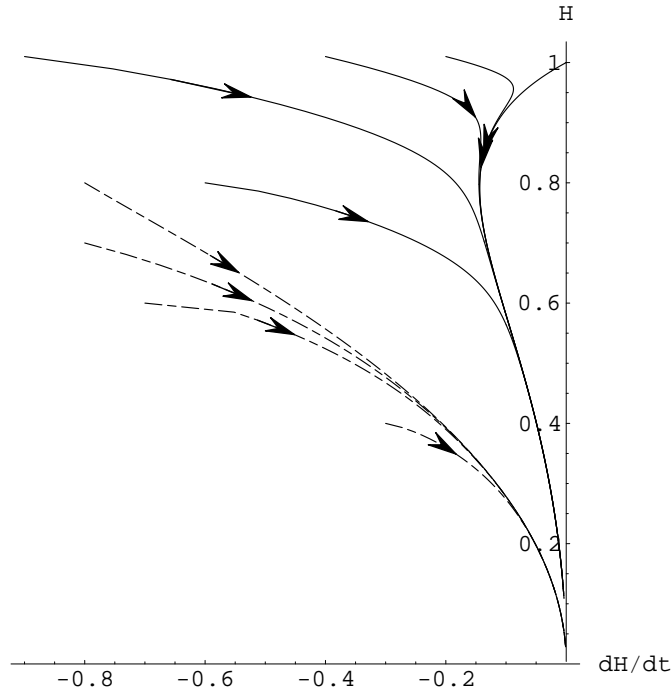}}   
\caption{Phase portrait for $f(R)=1/\ricci$. The unstable de Sitter solution is at the point $(0,1)$. The two late-time power-law
attractors are clearly visible for $p \simeq .77$ (dashed lines) and $p \simeq 3.22$ (solid lines). \label{fig3}}
\end{figure}

Consider a modification of the form $f(P)=-\mu^{6}/P$. The trace equation (\ref{constcurvexp}), indicates the de Sitter
constant curvature solution occurs for $R^{(P)}_{dS}= (16)^{1/3} \mu^{2}$. 
\newpage
The modified Friedmann equation takes the form
\bea
\frac{1}{8(3H^4 +3H^2\dot H + \dot H^2)^3}&=&(24 H^2\dot H^6 + 216 H^4\dot H^5+\m^6\dot H^4 
+ 864 H^6 \dot H^4  \nonumber \\ 
& +& 11\m^6 H^2 \dot H^3 + 1944 H^8 \dot H^3  + 2\m^6 H \dot H^2\ddot H  \nonumber \\ 
&+& 33 \m^6 H^4 \dot H^2+ 2592 H^{10} \dot H^2 
 + 30\m^6 H^6 \dot H \nonumber  \\ 
  &+& 6 \m^6 H^3 \dot H \ddot H + 1944 H^{12} \dot H + 6\m^6 H^8  \nonumber  \\ 
  &+& 648 H^{14} +  4\m^6 H^5\ddot H) =0   
 \,.
\eea
and power-law attractors are identified by substituting a power-law ansatz $H(t)=p/t$ and taking the late-time limit. By defining $v=-\dot H/H$
and requiring $v$ to be a constant $v_{0}=p$ we find the condition
\begin{equation}
6v_0^4-30v_0^3+41v_0^2-23v_0+5=0\ .
\end{equation}
which gives two late-time power-law attractors with $p=2\pm \sqrt{6}/2$. One of these corresponds to an accelerator with $p\simeq 3.22$,
while the other is not an accelerating solution with $p\simeq .77$. Both attractor solutions along with the de Sitter solution are easily identified
in the phase space portrait given in Fig.~(\ref{fig3}).
\subsection{Inverse powers of $Q \equiv \riemann$}
In the case of $f(Q)=-\mu^{6}/Q$ we find from the trace equation that the constant curvature de Sitter solution is given by
$R^{(Q)}_{dS}= (24)^{1/3}\mu^{2}$. The modified Friedmann equation is
\bea
\frac{1}{24[\dot H^2 + 2 H^2 \dot H + 2 H^4]^3}  \left[576 H^{14} + 1728 H^{12} \dot H  +2592 H^{10} \dot H^2 +
2304 H^8 \dot H^3 \right. \nonumber\\
+ 1296 H^6 \dot H^4 
+ 432 H^4 \dot H^5 +
72 H^2 \dot H^6 + 8\mu^6 H^8 + 36 \mu^6 H^6 \dot H +54 \mu^6 H^4 \dot H^2 \nonumber  \\
+ 20\mu^6 H^2 \dot H^3  \left. + 3\mu^6 \dot H^4
+ 4 \mu^6 H^5 \ddot H + 12 \mu^6 H^3 \dot H \ddot H + 
6\mu^6 H \dot H^2 \ddot H\right] = 0\ ,
\eea
which does not admit \emph{any} late-time power-law attractors.
\section{Conclusions}
We have described how a late-time period of cosmic acceleration emerges naturally in certain modified gravitational models. The models
in \cite{Carroll:2004de} involve inverse powers of linear combinations of curvature invariants.
Clearly, the modified gravitational models described in these notes are only a small handful of a much richer set of possible theories.
The late-time accelerating vacuum solutions we have discussed are not severely affected by the inclusion of matter. A simple way to 
understand this is the following. In an expanding universe the curvature invariants $R^{n}$, $P$ and $Q$ are all decreasing with time.
Therefore, any modifications of Einstein gravity that are inverse powers of such invariants are growing with time. If the universe
is initially matter dominated, these growing correction terms will eventually come to dominate the action.~\footnote{For a 
mathematically rigorous proof of this statement, see~\cite{Carroll:2004de}.}

It is an intriguing possibility that the observed cosmic acceleration of our universe may be a consequence of
some low-curvature modified gravitational theory. A first attempt to explore this possibility was made in~\cite{Carroll:2004de}. 
It is important that we try to better understand these 
models and the observational constaints on low energy modifications of GR. 
\section*{Acknowledgments}
I am grateful to my collaborators, Sean Carroll, Antonio De~Felice, Vikram Duvvuri, Mark Trodden and Michael Turner, and to
Renata Kallosh, John Moffat, Glenn Starkman and Alexei Starobinsky for helpful discussions and communications.
This work is supported in part by the National Science Foundation under grants PHY-0094122 and PHY-0354990, 
and by funds from Syracuse University.


\end{document}